\documentclass{optica-article}

\journal{opticajournal} 

\articletype{Research Article}

\usepackage{lineno}
\usepackage{booktabs}
\begin{document}

\title{The Taguchi method for optimizing nonlinear pulse propagation in optical fibers}

\author{Adity,\authormark{1,}\footnote[2]{Present address - Center for Nano Science and Engineering (CeNSE), Indian Institute of Science, Bangalore-560012.} and Srikanth Sugavanam,\authormark{1,*}}

\address{\authormark{1} Pratibimb Photonics Labs, School of Computing and Electrical Engineering, IIT Mandi, Kamand, Himachal Pradesh - 175005, India\\}

\email{\authormark{*}ssrikanth@iitmandi.ac.in}

\begin{abstract*} 
Understanding the nuances of nonlinear pulse propagation in optical fibers has led to several impactful applications across domains like optical communications, sensing and biophotonics. A key aspect in this regard is the use of appropriate optimization strategies for attaining requisite performance parameters. In this paper, we present the Taguchi method as a viable tool for optimizing nonlinear pulse propagation in optical fibers. We show that its use of the orthogonal arrays leads to rapid convergences to the desired pulse parameters, with even faster convergences obtained by favouring exploitation over exploration. We demonstrate the application of the method using two well-known problems from the field – the guiding center soliton, and soliton order conservation in dispersion decreasing fibers – which serve to underscore its salient features and also its potential for solution discovery across nonlinear pulse propagation problems. 

\end{abstract*}

\section{Introduction}

Pulse propagation in optical fibers is a multiparameter nonlinear problem, wherein the interaction of the light pulses with the fiber medium is mediated by the pulse properties, material and fiber dispersion, material nonlinearity, and also interaction lengths. The complexity of interactions is further enhanced by nonlinear spectral mixing and scattering interactions, which together with noise can lead to a wide diversity of operational regimes, ranging from self-organized behaviour on one end to chaotic behaviour on the other. Prime examples of such complex behaviour are the phenomena of formation of breathers and rogue waves \cite{Dudley2014, Akhmediev2021}, supercontinuum generation \cite{Jose2025}, or indeed, fiber lasers \cite{Woodward2018}. The complexity of interactions are often revealed in analytical or numerical investigations, including techniques like perturbative methods \cite{Malomed2002}. But while such methods offer interesting insights, their applicability to real-world systems, where higher-order interactions cannot be neglected becomes mathematically challenging. 
Furthermore, the higher dimensional nature of the parameter space also makes it challenging to explore, calling in for advanced investigative approaches. 

Indeed, in recent times there has been a proliferation of the use of optimization techniques or learning-based approaches with the increase in available compute capabilities, facilitating solution discovery over such highly multidimensional parameter spaces. For instance, techniques like genetic and particle swarm optimization have been used for optimizing performance of optical telecommunication links \cite{SimranjitSingh2014,Jiang2012, Redyuk2025},  finding suitable operational parameters for fiber lasers \cite{Lapre2023} and microcombs \cite{Mazoukh2024}. Similarly, there is increasing interest in the use of artificial intelligence (AI) and machine-learning based approaches for both simulation and solution discovery \cite{Sui2022,Sui2023}. Yet, there remain challenges from the point of view of compute time, compute capability, solution convergence, and explainability. AI and ML-based approaches require the generation of datasets that are sufficiently diverse to train the underlying learning architecture, which put increasing demands on compute and also storage. Likewise, while optimization techniques do not necessarily require dataset creation or storage, global search strategies like those employed by genetic algorithms can lead to long convergence times. Furthermore, many such optimization techniques are heuristic by nature, essentially meaning that the routine will converge to slightly different values over runs, even with the same initial seeding parameters. Above all, there is growing concern over the increasing use of data- and compute-intensive methods for such activities, as it has a direct impact on the environment \cite{Programme2024}. There is thus a growing need for finding memory and compute efficient techniques that can aid in effective exploration of such highly nonlinear and multidimensional parameter spaces for search of favourable solutions, and potentially solution discovery. 

In this paper, we show the Taguchi method as a viable optimization approach for the study of nonlinear pulse propagation in optical fibres. First proposed by Genichi Taguchi in the 1980s as a tool for industrial process optimization \cite{Taguchi1989}, the Taguchi method has its roots in the theory of the statistical Design of Experiments (DoE). The power of the technique lies in its use of orthogonal arrays in substantially reducing the number of experimental runs required to ascertain the contribution of the different experimental factors \cite{Kacker1991}.  
Within the photonics domain, the Taguchi method has been extensively used in laser machining \cite{Canel2012,Sheshadri2021} and optical sensor design \cite{Chen2004,Nan2022}, with modifications of the technique also being used for design optimization of antennas \cite{Weng2007}, and in design of optical communication links \cite{MAZ2016}. However insofar, the applicability of this technique has not been demonstrated for nonlinear pulse propagation in optical fibers. 

In the following, we first establish the foundations of the Taguchi method, followed by its implementation as an optimization technique. Next, we show how a pulse propagation problem can be recast as a Taguchi problem using two well-known problems in nonlinear pulse propagation – the guiding center soliton, and soliton propagation in a dispersion decreasing fiber. We then discuss the salient features of the Taguchi method, its scope, and its potential for use in such nonlinear pulse propagation problems. 

\section{The Taguchi Method}
In the parlance of the Taguchi framework, the response, i.e. the outcome of any experiment is determined by the interplay of the experimental variables, defined as factors. These factors can take different magnitudes, defined as levels, which can either be of a continuous or discrete nature. Ideally, it is desirable to perform a full factorial experiment, i.e. to verify the response of the experiment for all possible level permutations of the factors. However such an approach can become intractable very quickly if the experiment involves an appreciably large number of factors and/or levels. 

The Design of Experiments (DoE) based Taguchi method \cite{Kacker1991} utilizes a fractional factorial approach, where only a subset of the full factorial experiments is performed. Core to this method lies in its use of orthogonal arrays (OA), which realize a balanced combination of the levels of the different factors. Specifically, each column of the array is balanced, i.e. all levels occur the same number of times. Further, OAs can also be designed to have pairs, triads, or higher combinations of rows to be balanced, such that the row-wise combinations across them are never repeated. This is referred to as the strength of the OA. Correspondingly, an $k$-factor, $s$-level orthogonal array that is pairwise balanced and comprises of $M$ rows is represented symbolically as $OA(M,k,s,2)$. Typically, 3-level OAs are chosen for defining a central mean level, which is also amenable to optimization routines. The design of OA tables for different combinations of factors, levels and strengths is based on the foundations of Galois theory \cite{Hedayat1999}, and tabulations of the same are available widely (see for example, \cite{Sloane}).

\begin{figure}[t]
  \centering
  \includegraphics[width=\linewidth]{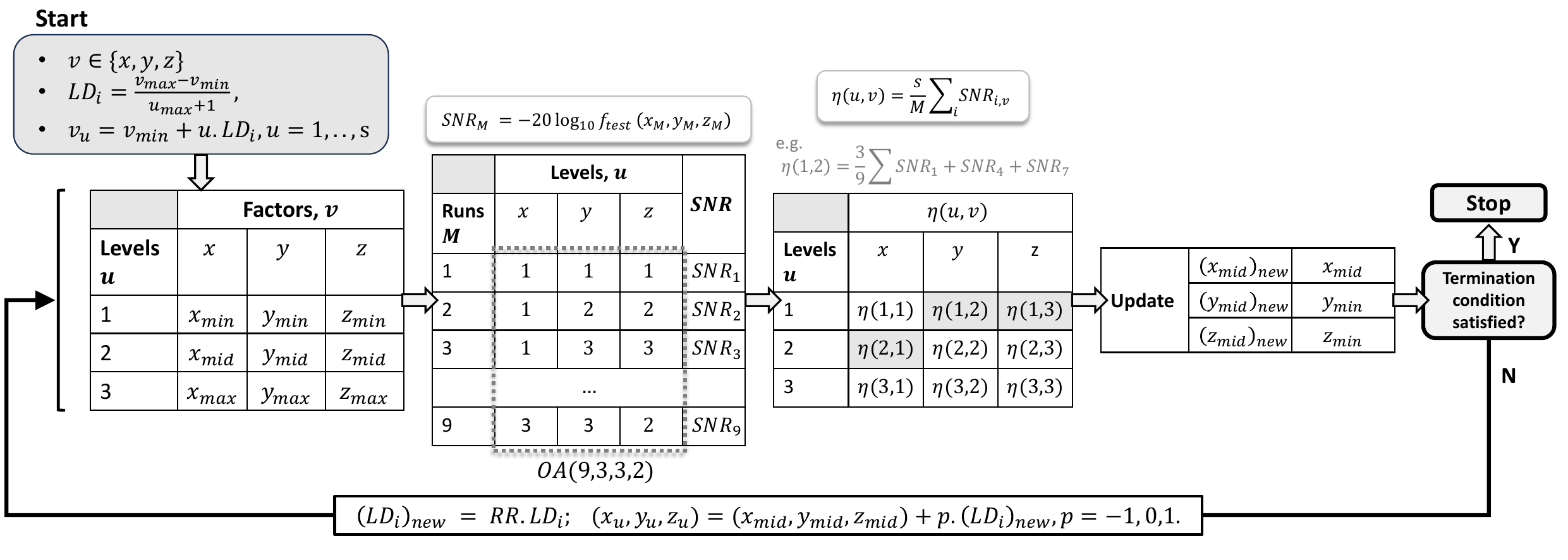}
\caption{Taguchi method workflow for a three-parameter three-level experiment.}
\label{Fig:1}
\end{figure}

Fig. \ref{Fig:1} shows the implementation workflow of the Taguchi method for a 3-factor, 3-level experiment, implemented using $OA(9,3,3,2)$. Each row of the OA represents an experimental run, which gives rise to a corresponding response. The maximum and minimum levels $v_{max}$ and $v_{min}$ of the factors are established based on \textit{a priori} knowledge of the experiment, which are then used to define the level differences $LD_i$. These are then used to initialize the $s$ levels of each the $k$ factors $v$, as defined by 
\begin{equation}
    v_u = v_{min}+u.LD_i, u = 1,...,s,
\end{equation}

where
\begin{equation}
    LD_i = \frac{v_{max}-v_{min}}{s+1}.
\end{equation}

The response of each experiment can be quantified by the response function $f_{test}(v)$. The Taguchi method defines responses as \textbf{N}-type (nominal the better), \textbf{S}-type (smaller the better), or \textbf{L}-type (larger the better) \cite{Ross1988}, which is decided on the basis of the nature of the process being monitored. This is analogous to defining an objective function in optimization problems, where an appropriate response function can be realized based on the problem and solution of interest. This is then converted to a log measure as per the Taguchi convention - 
\begin{equation}
SNR_M=-20\log_{10} f_{test}(v_M).    
\end{equation}
$f_{test}$ is conventionally defined to yield small values for the targeted response, thus yielding large values for the SNR as defined above. The response table $\eta(u,v)$ is then constructed by averaging the SNR values across all rows of the OA that contain the factor $v$ at the level $u$. The flow diagram shows an example of such a calculation for $\eta(1,2)$. As the rows of the OA are balanced, the contribution of a specific level $u$ of a given factor $v$ can be obtained by simply averaging all the rows that contain that specific factor at that specific level. Thus, the level $u$ for a given factor $v$ that leads to the largest value of $\eta(u,v)$ is then taken to be the level of the corresponding factor $v$ that leads to the desired optimal response.

For the problem of nonlinear pulse propagation, we employ a modification of the Taguchi method as described above, where it is transformed into an optimization routine \cite{Weng2007}.  
Essentially, after the first set of experimental runs are completed, the level that gives rise to the maximum response in the response table is chosen to be the new central level. Higher and lower levels are defined about this new central level employing the modified level difference factor $LD_i\times RR_i$, where $RR_i$ is called the reduction rate. $RR_i$ is chosen to be smaller than one, such that the parameter space of search reduces over subsequent iterations. The process is iterated multiple times till either the desired response (\textbf{L}-, \textbf{S}-, or \textbf{N}-type) is reached, or the difference in responses between consecutive iterations are less than a predefined tolerance. As we shall show below, the definition of the response thus becomes crucial for an effective exploration of the parameter space, while the reduction rate determines the extent of exploration and rate of convergence to the desired optimal solution.

\section{Results}

\subsection{The Guiding Center Soliton Problem}

 \begin{figure}[t]
  \centering
  \includegraphics[width=\linewidth]{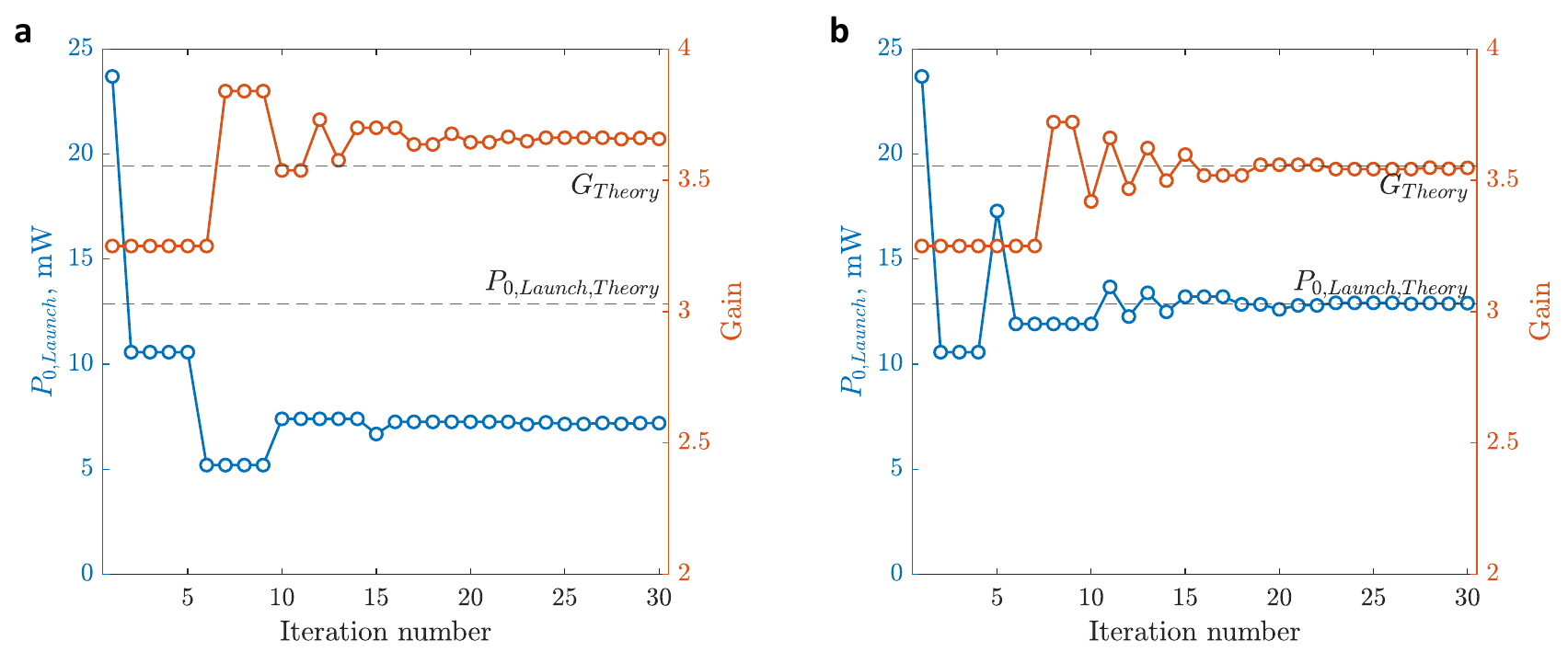}
\caption{Convergence results obtained with the Taguchi method for the guiding center problem with $RR = 0.8$ for a. $N_0 = 1$, b. $N_0 = 1.3285$.}
\label{Fig:2}
\end{figure}

For illustrating the use of the Taguchi method for optimizing nonlinear pulse propagation, we first consider the case of the guiding center soliton. In optical fibers, the soliton is a self-organized nonlinear optical pulse feature which preserves its shape as mediated by a balance of dispersion and nonlinearity. These properties had made them ideal candidates for optical communications, yet real-world fiber losses limited their applicability, requiring either distributed or periodic amplification schemes. The guiding center soliton approach \cite{Hasegawa1990,Blow1991} was the result of such studies, where it was shown that under specific launch power and periodic amplification conditions, an average solution for the shape-preserving soliton can be realized, leading to a periodic regeneration of the soliton just after each amplifier. Specifically, it was shown that for a given periodic amplifier spacing $L_A$ ($\ll$ dispersion length $L_D$), a soliton with order $N\approx 1$ could be sustained by choosing the gain $G_{Theory}$ and the launch soliton peak power $P_{0,Launch,Theory}$ as 
\begin{equation}\label{Eq:GS_Theo}
G_{Theory}=\exp{(\alpha L_A)}, 
P_{0,Launch,Theory}=\frac{G_{Theory}\ln G_{Theory}}{G_{Theory}-1}P_0,
\end{equation}
where $P_0$ is the peak power of the first order soliton for the given fiber. Eq. \ref{Eq:GS_Theo} shows that the gain essentially compensates the losses incurred by the soliton over each fiber span, with the launch power $P_{0,Launch,Theory}$ being a direct function of the gain. 

The relation between the gain and soliton launch power was originally established using a perturbative approach. To apply the Taguchi method, we reformulate the problem, starting with the assumption that there is no relation between the gain of the amplifiers $G$ and the soliton launch power $P_{0,Launch}$. That is, we consider $G$ and $P_{0,Launch}$ as experimental factors, with the objective of finding their appropriate combination which will yield to a periodically regenerating soliton after each amplifier. The outcome of the experiment is then ascertained by simulating propagation of the launched soliton over a configuration of $M_{amp}$ periodically spaced amplifiers. Towards this, an adaptive split-step Fourier transform routine is implemented for the corresponding experimental factors with a phase tolerance of $10^{-5}$. A sech-pulse of intensity FWHM 50 ps is considered as the launch pulse. The fiber has a group velocity dispersion (GVD) of -7650 $fs^2/m$, with loss $\alpha$ = 0.2611 dB/km and nonlinearity $\gamma$ = 1.3 $W^{-1} km^{-1}$. The amplifier spacing is taken as $L_A = 0.2\times L_D$, where $L_D = 105.5$ km. Propagation is considered over a total distance of 200 km. As per Eq. \ref{Eq:GS_Theo}, for these propagation parameters $G_{Theory} = 3.5556, P_{0,Launch,Theory}=12.87 $ mW. 

The Taguchi response function is then defined as   

\begin{equation}\label{Eq:f_GS}
f_{test} = \sqrt{\frac{1}{M_{amp}}\left[\sum_{i=1}^{M_{amp}}{\left(N_0-N_i\right)^2}\right]+\left(\frac{P_{0,M_{amp}}-P_{0,Launch}}{P_{0,Launch}}\right)^2 +\left(\frac{T_{0,M_{amp}}-T_{0,Launch}}{T_{0,Launch}}\right)^2},    
\end{equation}

where $N_i$ refer to the soliton order after just after each amplifier, while ($P_{0,Launch}$,$T_{0,Launch}$) and ($P_{0,M_{Amp}}$,$T_{0,M_{Amp}}$) are the peak powers and pulse widths of the soliton at launch and at the final amplifer $M_{Amp}$ respectively. The response function of Eq. \ref{Eq:f_GS} is defined in the $\textbf{N}$-form, and searches for a soliton of order $N_0$. The remaining two terms of the function avoid convergence to trivial solutions (e.g. $P_{0,Launch}=0, G = 0$). Presently, $N_0$ is set as 1 for seeking propagating solutions close to a first order soliton.
\begin{figure}[t]
  \centering
  \includegraphics[width=\linewidth]{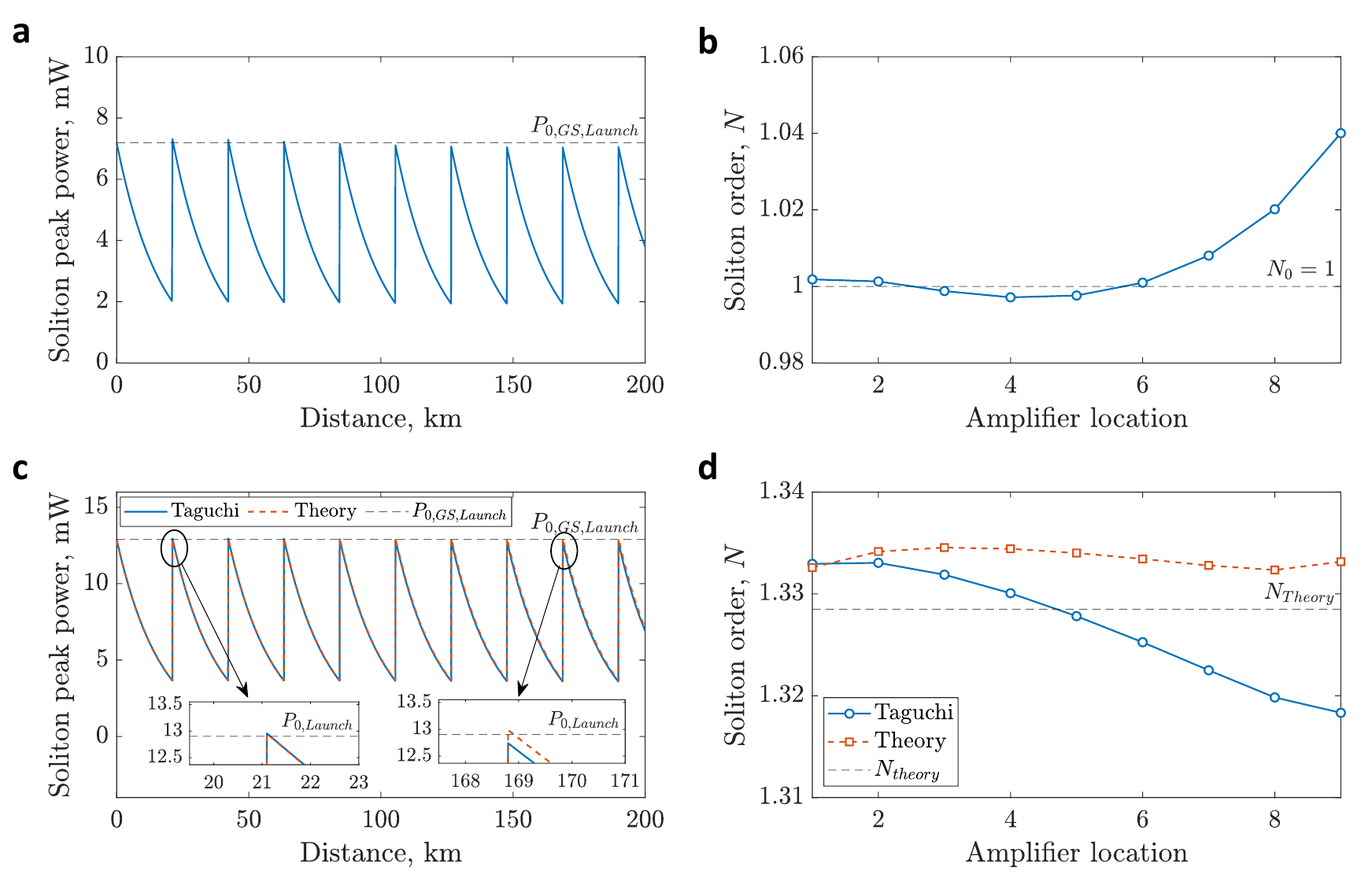}
\caption{Propagation characteristics of the guiding center soliton obtained with the Taguchi method for a, b. $N_0 = 1$; c, d. $N_0 = 1.3285$. The soliton orders after each amplifier location (Figs. b,d) are close to the desired value $N_0$. The orange dotted plots (Figs.c,d) and insets show the results of the split-step method obtained using the theoretical values of the gain and launch power for comparison.}
\label{Fig:2b}
\end{figure}

While it is possible to consider the gain $G$ and peak power of the launched soliton $P_{0,Launch}$ directly as the Taguchi factors,  a normalization can be applied to keep uniformity of the order of magnitudes of the factors. Here, we normalize the peak power $P_{0,Launch}$ with the peak power of an $N=1$ soliton for the present configuration, giving rise to the normalized factor $P_{fac}$ that is used as a Taguchi factor. This is in alignment with the expectation that the periodic amplifier configuration will support the propagation of an $N\approx 1$ soliton. In addition, this allows the initial parameter search space to be set at [1, 10] for both the factors $G$ and $P_{fac}$. Note that this search space is sufficiently large in comparison with the theoretically expected values, allowing for sufficient exploration of the parameter space and a robust test of the Taguchi method. 
The orthogonal array $OA(9,2,3,2)$ is used for carrying out the experiments. After the completion of the experimental runs, a response table is generated by collating the average SNR of each level of the experimental factor, which enables the determination of the new central levels. Incidentally the orthogonal array for a 2-factor, 3-level experiment results in a full-factorial experiment. While this is not favourable from an experimental design perspective, the Taguchi optimization as represented in Fig. \ref{Fig:1} still remains applicable, with the chosen example revealing several of its important aspects. 

\begin{figure}[t]
  \centering
  \includegraphics[width=\linewidth]{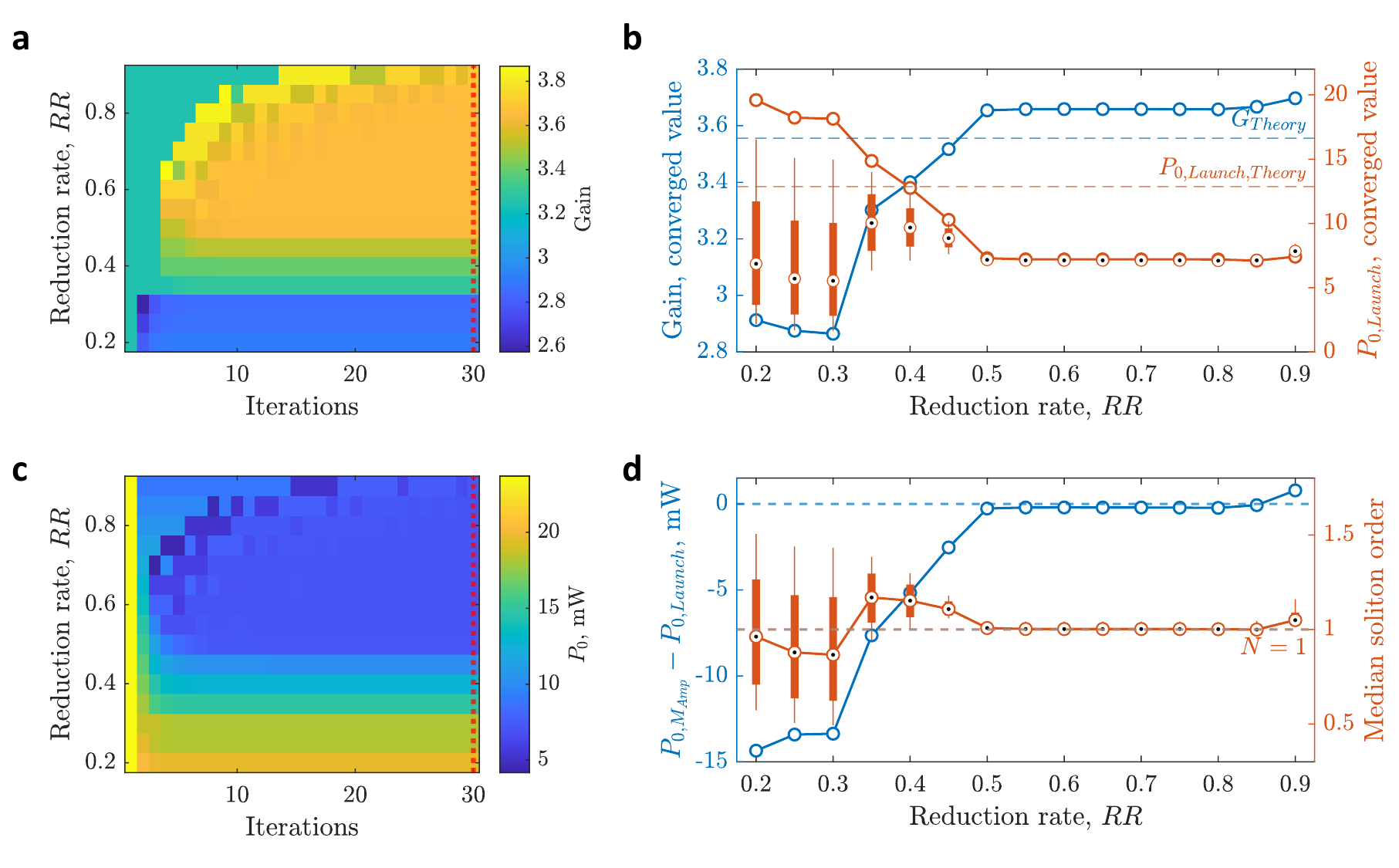}
\caption{Exploration vs. exploitation control using the reduction rate $RR$ with $N_0 = 1$. Convergence characteristics of a. Gain $G$ and c. Soliton power $P_{0,Launch}$ across iterations with the reduction rate as a parameter; b. Converged $G$ and $P_{0,Launch}$ values after 30 iterations; d. Difference between launched soliton power and that from the last amplifier (left axis), and median soliton order across amplifers (right axis). Box plots (Figs. b, d) show the distribution of the peak soliton powers and the soliton order just after each amplifier in the chain. Low variation in the peak power and soliton order characteristics across a range of $RR$ values confirm the guided nature of the solution. 
}
\label{Fig:3}
\end{figure}

Fig. \ref{Fig:2}a show the results obtained for the response function as indicated in Eq. \ref{Eq:f_GS} by setting $N_0 = 1$ for a reduction rate $RR = 0.8$, where $P_{0,Launch}=P_0\times P_{fac}$. The Taguchi method based optimization routine attains convergence in about 20 iterations, corresponding to a total of 180 experimental runs. This observed convergence is faster than commonly used optimization algorithms such as GA and PSO, an observation which has also been reported elsewhere (e.g. \cite{MAZ2016}, also see Supplementary information). The speed-up can be attributed to the DoE-based approach of the Taguchi method, in contrast to the randomized initial seeding conditions of GA and PSO. Interestingly, the convergence of the factors $G$ and $P_{0,Launch}$ is away from the theoretically predicted results. This is because the function $f_{test}$ seeks an $N=1$ solution, while the theory posits the launching of an $N = 1.3285$ order soliton. Indeed, upon setting $N_0=1.3285$ in Eq. \ref{Eq:f_GS} leads to convergence to the theoretical values as shown in Fig. \ref{Fig:2}b. This shows that the Taguchi method can become a tool for solution discovery, in this case with a potential for tuning the soliton order across the dimensions of the gain $G$ and the soliton launch peak power $P_{0, Launch}$. Further verification of the nature of convergence is shown in Fig. \ref{Fig:2b}, obtained by using the adaptive split-step routine to propagate the guiding center soliton with the converged values for gain and soliton launch power for the chosen values of $N_0 = 1$ and $N_0 = 1.3285$  in Fig. \ref{Fig:2}. Figs. \ref{Fig:2b}a, c show the periodic regeneration of the soliton after each amplifier location as expected for a guiding center configuration, while Figs. \ref{Fig:2b}b, d show the soliton orders after each amplifier for the two cases. The results obtained for $N_0 = 1.3285$, show that the Taguchi method converges sufficiently close to the expected theoretical values in keeping with real-world precisions and tolerances. The precision of convergence can be improved by choosing a higher value for the phase tolerance (e.g. $10^{-6},10^{-8}$), or by setting a tolerance-based termination condition for the Taguchi routine.

The reduction rate $RR$ plays an important role in controlling the trade-off between exploration and exploitation. A larger reduction rate ($RR\lesssim1$) implies a slower reduction in the size of factor space over iterations, leading to a larger search space for the optimization routine. Keeping a larger value of the reduction rate thus promotes exploration and the discovery of global extrema. However, this also leads to longer convergence times. A smaller reduction rate can enable faster convergence but they may lead to local solutions, or solution stagnation. Fig. \ref{Fig:3} reveal the exploration vs. exploitation characteristics obtained by varying $RR$ for $N_0=1$. Figs. \ref{Fig:3}a,c show how moving to a smaller value of $RR$ leads to a faster convergence. Fig. \ref{Fig:3}b shows the results of convergence (red dotted lines in Figs. \ref{Fig:3}a, c) to the gain $G$ of the amplifiers and the soliton launch power $P_{0,GS}$ obtained across the different reduction rates. Here the orange box plot captures the extent of variation of the soliton peak power just after each amplifier. This shows that the Taguchi method shows a convergence to an $N=1$ soliton solution for the guiding center soliton problem over a wide range of reduction rates, even for reduction rates as small as $RR=0.5$ in as low as 8 iterations (72 split-step simulations). Fig. \ref{Fig:3}d additionally evidences the nature of convergences, with the blue plot (left axis) showing the difference in peak powers between the launched soliton and that from the last amplifier in the chain, and the orange plot (right axis) showing the median soliton order across the amplifiers being close to $N=1$ across the amplifiers. The box plot overlay shows the distribution of the soliton orders across the amplifiers for the converged values, which again shows that the converged solutions exhibit a regeneration of a first order soliton across all amplifiers. The results show that the Taguchi method can indeed aid solution discovery beyond those predicted by theory, and are amenable to experimental verification. 

\subsection{Dispersion decreasing fibers}
The guiding center soliton problem above example highlights the fast convergence and exploration-exploitation trade-off control of the Taguchi based optimization approach. However, as it is only a 3-level, 2-parameter problem, the resulting OA leads to a full factorial experimental design. We now show the applicability of a fractional factorial experimental design to the case of soliton propagation in dispersion decreasing fibers \cite{Tajima1987,Richardson1995}. In such fibers, the exponentially decreasing dispersion along the length of the fiber balances the loss-mediated reduction of self-phase modulation along the length of the fiber span. The theoretically ascertained dispersion profile is given as 

\begin{equation}\label{Eq:DDF_Theory}
    |\beta_2(z)|=|\beta_{2,0}|\exp(-\alpha z),
\end{equation}

where $\beta_{2,0}$ is the GVD of the fiber at distance $z=0$, i.e. the launch end of the fiber, and $\alpha$ is the fiber loss parameter. 

For using the Taguchi method to arrive at a suitable dispersion profile that conserves the soliton order, we reformulate the problem as finding the Taylor series coefficients of the dispersion profile, correct up to the third order in distance. Specifically, the group velocity dispersion $\beta_2$ of the fiber is modelled as a Taylor series expansion of the form

\begin{equation}\label{Eq:DDF_Taylor}
\beta_2(z)= \beta_{2,0}\left[1+C_0a_1z+\frac{(C_0a_2)^2}{2!}z^2+\frac{(C_0a_3)^3}{3!}z^3 \right],
\end{equation}

where $a_i$ give a measure of the magnitude of the coefficients, while $C_0$ takes the sign of the term into account. Correspondingly, the response function can be modelled to look for first order soliton solutions. Two additional steps are carried out to facilitate the implementation of the optimization routine. Firstly, given the power series nature of Eq. \ref{Eq:DDF_Taylor}, the coefficients $a_i$ can be expected to differ by orders of magnitude. To avoid this, instead of $a_i$ the logarithmically scaled coefficients $A_i = \log_{10}(a_i)$ are taken to be the experimental factors. Secondly, the order of the soliton $N$ is measured at periodic intervals $M_{loc}$ along the span of the dispersion-tailored fiber to enable a definition of the response function of the form 

\begin{equation}\label{Eq:DDF_resp}
f_{test} = \sqrt{\frac{1}{M_{loc}}\sum_{i=1}^{M_{loc}}(1 - N_i)^2 }
\end{equation}

\begin{figure}[t]
  \centering
  \includegraphics[width=\linewidth]{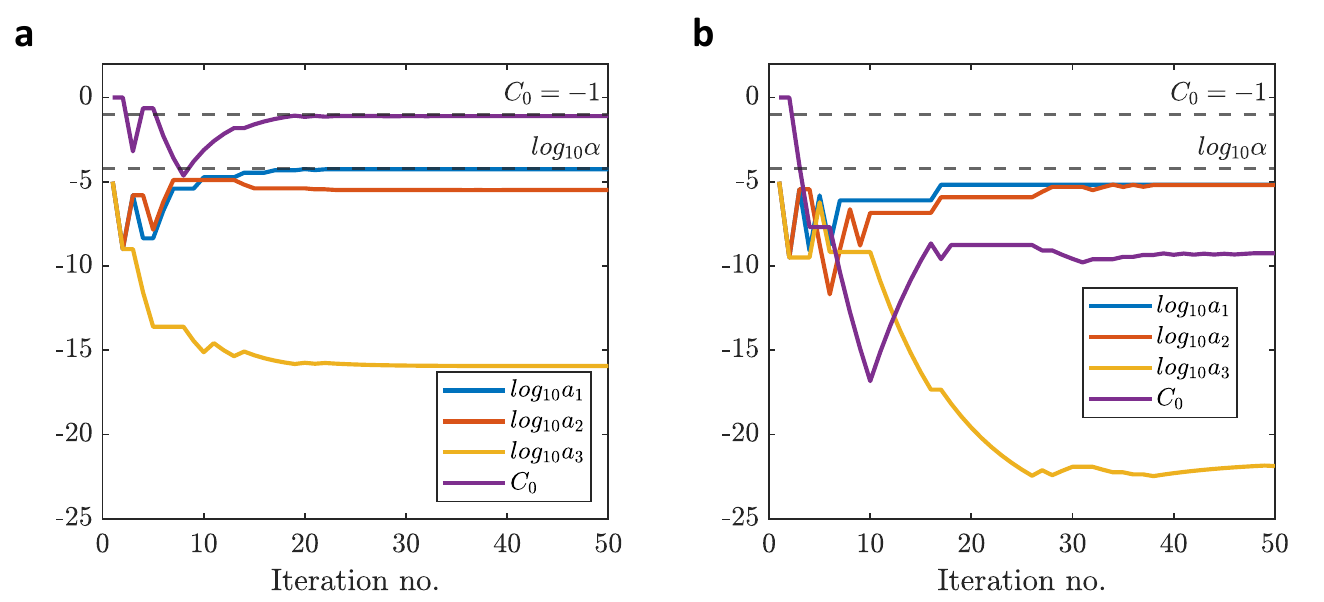}
\caption{Convergence characteristics obtained with the Taguchi method for the dispersion decreasing fiber problem for two reduction rates; a. $RR = 0.8$, b. $RR=0.9$.}
\label{Fig:4}
\end{figure}

\begin{figure}[h]
  \centering
  \includegraphics[width=\linewidth]{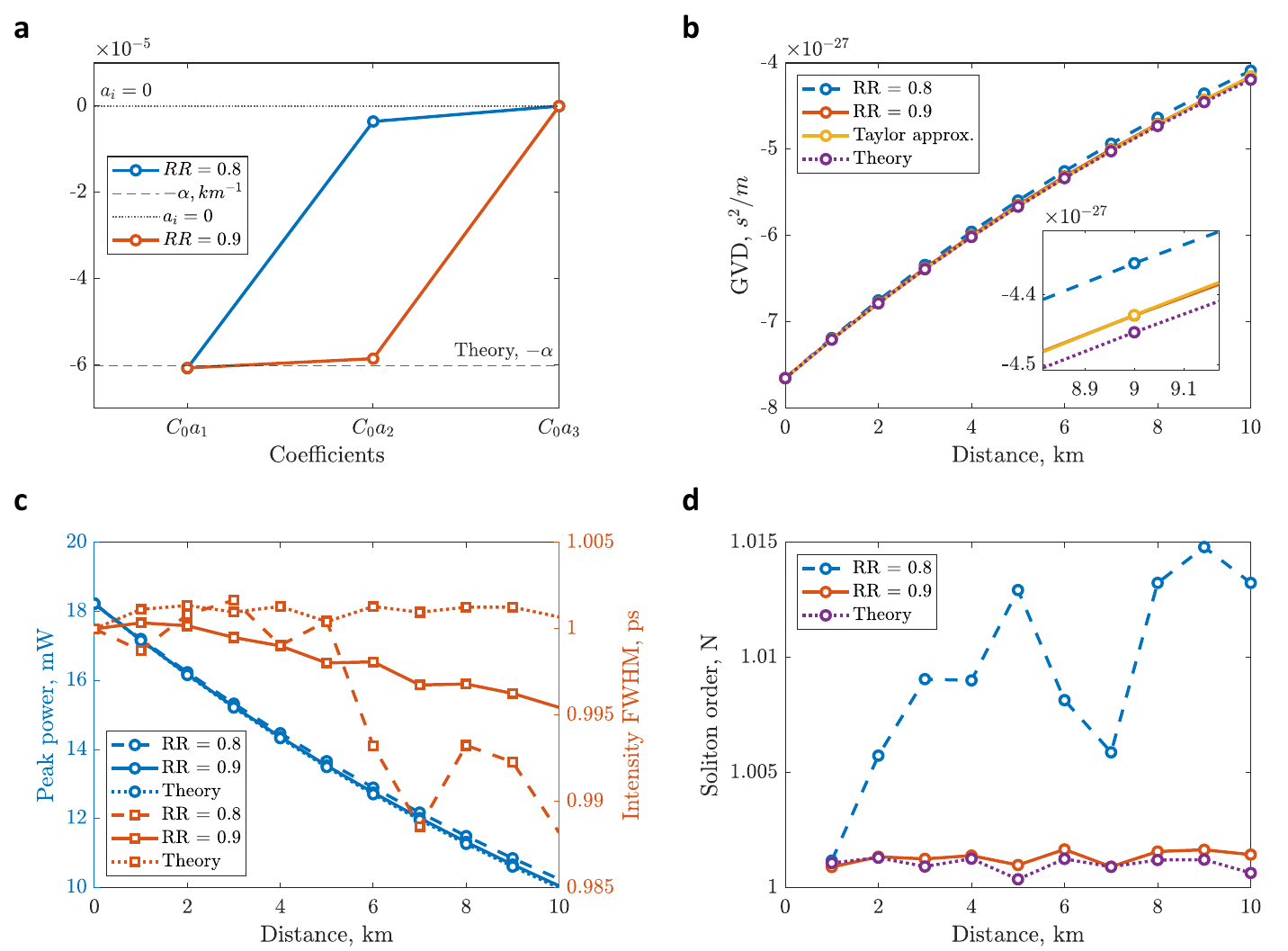}
\caption{Verification of convergence for the dispersion decreasing fiber problem. a. Convergence of the product terms $C_0a_i$ for $RR=0.8$ and $RR = 0.9$;  b. Resulting dispersion profiles, with inset showing good agreement between theory and the Taguchi method; c. Results obtained by propagation of converged solution, comparing soliton peak power and pulse width; d. Soliton order variation, with the results obtained using the Taguchi method with $RR=0.9$ approaching those obtained with the theoretical dispersion profile of Eq. \ref{Eq:DDF_Theory}.}
\label{Fig:5}
\end{figure}

Thus the Taguchi problem is to find the values of the experimental factors - the logarithmically scaled coefficients $A_1, A_2,$ and $A_3$, and the coefficient $C_0$ - corresponding to the constraint set by the response function of Eq. \ref{Eq:DDF_resp}. Here, a strength-2 orthogonal array of the form $OA(9,4,3,2)$ is chosen for the problem, which results in 9 experimental runs for orthogonal permutations of the 4 factors across 3 levels. This thus results in a fractional factorial experiment, leading to a 9-fold decrease in experimental runs. 

The physical problem of the soliton propagation in the dispersion tailored fiber is modeled 
considering the propagation of 1 ps sech- solitons over a distance of 10 km. The dispersion at the beginning of the fiber, $\beta_{2,0}$, is taken to be -7650 $fs^2/m$. The nonlinearity is taken to be 1.3 $W^{-1} km^{-1}$. An adaptive split step routine with a phase tolerance of $10^{-6}$ is used for modelling the propagation. The range for both $A_i$ and $C_0$ are set to be [-10, 10]. The results of convergence are investigated for two reduction rates, $RR = 0.8, 0.9$. The number of iterations is limited to 50, resulting in a total of 450 split-step NLSE runs. In addition, to aid faster convergence, the first four iterations are implemented using a conventional fixed discrete step split-step routine, as starting with the adaptive-step routine will lead to exceptionally large execution times owing to the initially large values of the experimental factors.

Fig. \ref{Fig:4} shows the results of the implementation of the Taguchi method for finding the dispersion profile of the fiber for the two reduction rates $RR = 0.8, 0.9$ and compares it with the theoretically known solution. Note that the convergence characteristics are shown for the logarithmically scaled values $A_i$. For $RR = 0.8$, convergence is observed for the coefficients $C_0$ and $A_1$ to the theoretically expected values, while the values for $A_2$ and $A_3$ do not. For $RR = 0.9$, interestingly it is seen that none of the coefficients $C_0$ or $A_i$ converge to the theoretically expected values. However, as seen in Fig. \ref{Fig:5}a, the corresponding factors $C_0a_i$ are closer to the theoretically expected values for the higher reduction rate. This is in agreement with our understanding that a larger reduction rate permits more exploration of the parameter space, which leads to a better estimation in line with the imposed constraint. Indeed, we see in Fig. \ref{Fig:5}b a very good agreement between the theoretically expected exponentially decreasing profile for the fiber with the results obtained with the Taguchi method. The Taylor series expansion corresponding to the theoretical exponential dispersion is also shown for comparison of the accuracy of the approximation, where a minor deviation is seen at large distances (see inset). 
The dispersion profiles as obtained with the Taguchi method show small deviations at large distances.
While these differences are small, Figs. \ref{Fig:5}c and d show even such small differences affect the evolution characteristics. For instance, the intensity FWHM of the pulse remains very close to the launched pulse width of 1 ps when considering the pulse evolution as per the theoretical exponential profile. In comparison, the pulse widths fall with distance for the dispersions profile obtained with the Taguchi method. Yet, the differences are well within experimental limits, with the pulse width being correct to within 10 fs for $RR = 0.9$ over the observed span of 10 km. Similarly, the soliton order is remarkably close to the theoretically expected value of $N = 1$, with the deviations predominantly arising from the difference in the deviation of the dispersion profile from the theoretical. This is seen in Fig. \ref{Fig:5}d where the soliton orders obtained for $RR=0.9$ is compared with those obtained with the theoretical dispersion profile of Eq. \ref{Eq:DDF_Theory}. As in the guiding center soliton case, the obtained results can be improved by increasing the reduction rate, by increasing the adaptive split-step phase tolerance, and also by increasing the number of iterations to allow for convergence with a higher precision. Also, while the convergence is not to the exact Taylor series coefficient values, the resulting dispersion profile is very close to that predicted by theory, allowing for obtaining closest-fit approximations that can lead to uncovering underlying physical mechanisms of the evolution of the pulses in such media. Additionally, the results indicate that over the investigated propagation distance a quadratic dispersion profile can sufficiently ensure conservation of the soliton order. Such insights can lead to simplification of the manufacturing process and experimental realization. 

The results above demonstrate the applicability of the fractional factorial design-based Taguchi approach for optimizing nonlinear pulse propagation. It also shows the importance of an appropriate choice of Taguchi factors and normalization of their relative magnitudes for promoting convergence.  

\section{Discussions}

The field of nonlinear fiber optics has contributed immensely to the world as we know it today, which is quite evident when we look at  areas like optical telecommunications, biophotonics and sensing. Beyond this, the fiber optic platform offers a remarkable and versatile laboratory for studying fundamental nonlinear physics. The Taguchi method demonstrated above adds a versatile tool to the arsenal of attacking tough nonlinear problems and motivating solution discovery. The primary advantages of the speed of convergence and the fractional factorial design can significantly cut down requirements on computational resources, which can have a direct positive impact on resource usage and carbon footprint. The guiding center soliton and the dispersion decreasing fiber problems were chosen for the present work as they are well-known and understood, and they help in underscoring the salient features of the Taguchi method. The principles and approaches shown here can be extended to other higher dimensional problems like fiber laser design, supercontinuum generation, or even optimization of frequency combs, with the help of higher dimensional orthogonal arrays. In fact the reduction in the number of experimental runs becomes more pronounced as the number of factors increase. The demonstrated examples also highlight the importance of parameter choice and objective function definition. Given the simplicity of its implementation, the Taguchi method can be first used to test the effectiveness of the designed objective functions before deploying them on more complex and compute-intensive optimization routines. While the Taguchi method allows for an exploration vs. exploitation trade-off via control of the reduction rate, convergence to the global extrema is not guaranteed. However, this problem is not unique to the Taguchi method, and is common across optimization methodologies exploring higher dimensional complex problems that require additional tests for verifying the nature of convergence. In this context, the Taguchi method can be deployed as a first step, followed by verification with larger reduction rates and using other optimization routines like the GA and PSO, which while time-intensive can be targeted to look for global solutions. 

In all, in light of its simplicity of implementation, the fractional factorial approach, and its fast convergence characteristics, the Taguchi method offers an alternative route for optimizing nonlinear pulse propagation in optical fibres. Furthermore, the well-established field of statistical Design of Experiments can offer further tools that can augment its primary advantages, leading to more robust designs which is the core tenet of the Taguchi method.

\begin{backmatter}
\bmsection{Funding}
The authors acknowledge the funding received from the DST-CRG Grant, as awarded by the Science and Engineering Research Board, a statutory body of the Department of Science and Technology (DST), Government of India. 

\bmsection{Acknowledgment}
The authors thank Dr. V. Balaswamy, Dr. Siddhartha Sarma and Mr. Iqbal Ashraf for insightful and fruitful discussions.  

\bmsection{Disclosures}
The authors declare no conflicts of interest.

\bmsection{Data availability} Data underlying the results presented in this paper are not publicly available at this time but may be obtained from the authors upon reasonable request.

\bibliographystyle{ieeetran}
\bibliography{ReferencesJR}

@Article{Akhmediev2021,
  author     = {Akhmediev, Nail},
  journal    = {Frontiers in Physics},
  title      = {Waves that {Appear} {From} {Nowhere}: {Complex} {Rogue} {Wave} {Structures} and {Their} {Elementary} {Particles}},
  year       = {2021},
  issn       = {2296-424X},
  month      = jan,
  note       = {45 citations (Crossref) [2025-05-05] Publisher: Frontiers},
  volume     = {8},
  doi        = {10.3389/fphy.2020.612318},
  language   = {English},
  shorttitle = {Waves that {Appear} {From} {Nowhere}},
  url        = {https://www.frontiersin.orghttps://www.frontiersin.org/journals/physics/articles/10.3389/fphy.2020.612318/full},
  urldate    = {2025-05-05},
}

@Article{Jose2025,
  author     = {Jose, Amala and Chowdhury, Sourav Das and Balasubramanian, Sudharsan and Krupa, Katarzyna and Wang, Zhiqiang and Upadhyay, B. N. and Grelu, Philippe and Kanagaraj, Nithyanandan},
  journal    = {Laser \& Photonics Reviews},
  title      = {Noise-{Like} {Pulse} {Seeded} {Supercontinuum} {Generation}: {An} {In}-{Depth} {Review} {For} {High}-{Energy} {Flat} {Broadband} {Sources}},
  year       = {2025},
  issn       = {1863-8899},
  note       = {1 citations (Crossref) [2025-05-05] \_eprint: https://onlinelibrary.wiley.com/doi/pdf/10.1002/lpor.202400511},
  number     = {5},
  pages      = {2400511},
  volume     = {19},
  copyright  = {© 2024 Wiley-VCH GmbH},
  doi        = {10.1002/lpor.202400511},
  language   = {en},
  shorttitle = {Noise-{Like} {Pulse} {Seeded} {Supercontinuum} {Generation}},
  url        = {https://onlinelibrary.wiley.com/doi/abs/10.1002/lpor.202400511},
  urldate    = {2025-05-05},
}

@Article{Dudley2014,
  author   = {Dudley, John M. and Dias, Frédéric and Erkintalo, Miro and Genty, Goëry},
  journal  = {Nature Photonics},
  title    = {Instabilities, breathers and rogue waves in optics},
  year     = {2014},
  issn     = {1749-4893},
  month    = oct,
  note     = {792 citations (Crossref) [2025-05-05]},
  number   = {10},
  pages    = {755--764},
  volume   = {8},
  doi      = {10.1038/nphoton.2014.220},
  url      = {https://doi.org/10.1038/nphoton.2014.220},
}

@Article{Woodward2018,
  author   = {Woodward, R I},
  journal  = {Journal of Optics},
  title    = {Dispersion engineering of mode-locked fibre lasers},
  year     = {2018},
  issn     = {2040-8986},
  month    = feb,
  note     = {79 citations (Crossref) [2025-05-05] Publisher: IOP Publishing},
  number   = {3},
  pages    = {033002},
  volume   = {20},
  doi      = {10.1088/2040-8986/aaa9f5},
  language = {en},
  url      = {https://dx.doi.org/10.1088/2040-8986/aaa9f5},
  urldate  = {2025-05-05},
}

@InCollection{Malomed2002,
  author    = {Malomed, Boris A.},
  booktitle = {Progress in {Optics}},
  publisher = {Elsevier},
  title     = {Chapter 2 - {Variational} methods in nonlinear fiber optics and related fields},
  year      = {2002},
  editor    = {Wolf, E.},
  month     = jan,
  pages     = {71--193},
  volume    = {43},
  doi       = {10.1016/S0079-6638(02)80026-9},
  url       = {https://www.sciencedirect.com/science/article/pii/S0079663802800269},
  urldate   = {2025-05-05},
}

@Article{SimranjitSingh2014,
  author  = {{Simranjit Singh} and {Sonak Saini} and {Gurpreet Kaur} and {Rajinder Singh Kaler}},
  journal = {Optical Engineering},
  title   = {Multiparameter optimization of a {Raman} fiber amplifier using a genetic algorithm for an {L}-band dense wavelength division multiplexed system},
  year    = {2014},
  month   = jan,
  note    = {10 citations (Crossref) [2025-05-05]},
  number  = {1},
  pages   = {016103},
  volume  = {53},
  doi     = {10.1117/1.OE.53.1.016103},
  url     = {https://doi.org/10.1117/1.OE.53.1.016103},
}

@Article{Jiang2012,
  author   = {Jiang, Hai Ming and Xie, Kang and Wang, Ya Fei},
  journal  = {Optics and Lasers in Engineering},
  title    = {Flat gain spectrum design of {Raman} fiber amplifiers based on particle swarm optimization and average power analysis technique},
  year     = {2012},
  issn     = {0143-8166},
  month    = feb,
  note     = {12 citations (Crossref) [2025-05-05]},
  number   = {2},
  pages    = {226--230},
  volume   = {50},
  doi      = {10.1016/j.optlaseng.2011.08.012},
  url      = {https://www.sciencedirect.com/science/article/pii/S0143816611002569},
  urldate  = {2025-05-05},
}

@Article{Redyuk2025,
  author   = {Redyuk, Alexey and Shevelev, Evgeny and Danilko, Vitaly and Fedoruk, Mikhail},
  journal  = {Journal of Lightwave Technology},
  title    = {{ML}-{Assisted} {Particle} {Swarm} {Optimization} of a {Perturbation}-{Based} {Model} for {Nonlinearity} {Compensation} in {Optical} {Transmission} {Systems}},
  year     = {2025},
  issn     = {1558-2213},
  month    = mar,
  note     = {1 citations (Crossref) [2025-05-05]},
  number   = {5},
  pages    = {2143--2150},
  volume   = {43},
  doi      = {10.1109/JLT.2024.3487204},
  url      = {https://ieeexplore.ieee.org/abstract/document/10737152},
  urldate  = {2025-05-05},
}

@Article{Sui2022,
  author    = {Sui, Hao and Zhu, Hongna and Luo, Bin and Taccheo, Stefano and Zou, Xihua and Yan, Lianshan},
  journal   = {Optics Letters},
  title     = {Physics-based deep learning for modeling nonlinear pulse propagation in optical fibers},
  year      = {2022},
  issn      = {1539-4794},
  month     = aug,
  note      = {15 citations (Crossref) [2025-05-05] Publisher: Optica Publishing Group},
  number    = {15},
  pages     = {3912--3915},
  volume    = {47},
  copyright = {© 2022 Optica Publishing Group},
  doi       = {10.1364/OL.460489},
  language  = {EN},
  url       = {https://opg.optica.org/ol/abstract.cfm?uri=ol-47-15-3912},
  urldate   = {2025-05-05},
}

@Article{Sui2023,
  author    = {Sui, Hao and Zhu, Hongna and Jia, Huanyu and Li, Qi and Ou, Mingyu and Luo, Bin and Zou, Xihua and Yan, Lianshan},
  journal   = {Optics Letters},
  title     = {Predicting nonlinear multi-pulse propagation in optical fibers via a lightweight convolutional neural network},
  year      = {2023},
  issn      = {1539-4794},
  month     = sep,
  note      = {5 citations (Crossref) [2025-05-05] Publisher: Optica Publishing Group},
  number    = {18},
  pages     = {4889--4892},
  volume    = {48},
  copyright = {© 2023 Optica Publishing Group},
  doi       = {10.1364/OL.496973},
  language  = {EN},
  url       = {https://opg.optica.org/ol/abstract.cfm?uri=ol-48-18-4889},
  urldate   = {2025-05-05},
}

@Book{Programme2024,
  author     = {Programme, United Nations Environment},
  title      = {Artificial {Intelligence} ({AI}) end-to-end: {The} {Environmental} {Impact} of the {Full} {AI} {Lifecycle} {Needs} to be {Comprehensively} {Assessed} - {Issue} {Note}},
  year       = {2024},
  isbn       = {978-92-807-4182-7},
  month      = sep,
  language   = {English},
  shorttitle = {Artificial {Intelligence} ({AI}) end-to-end},
  url        = {https://wedocs.unep.org/xmlui/handle/20.500.11822/46288},
  urldate    = {2025-05-05},
}

@Article{Kacker1991,
  author   = {Kacker, R.N. and Lagergren, E.S. and Filliben, J.J.},
  journal  = {Journal of Research of the National Institute of Standards and Technology},
  title    = {Taguchi's orthogonal arrays are classical designs of experiments},
  year     = {1991},
  issn     = {1044-677X},
  month    = sep,
  note     = {173 citations (Crossref) [2025-05-05]},
  number   = {5},
  pages    = {577},
  volume   = {96},
  doi      = {10.6028/jres.096.034},
  language = {en},
  url      = {https://nvlpubs.nist.gov/nistpubs/jres/096/jresv96n5p577_A1b.pdf},
  urldate  = {2025-05-05},
}

@Article{Chen2004,
  author   = {Chen, Xiaopei and Zhang, Yan and Pickrell, Gary and Antony, Jiju},
  journal  = {International Journal of Productivity and Performance Management},
  title    = {Experimental design in fiber optic sensor development},
  year     = {2004},
  issn     = {1741-0401},
  month    = jan,
  note     = {7 citations (Crossref) [2025-05-05] Publisher: Emerald Group Publishing Limited},
  number   = {8},
  pages    = {713--725},
  volume   = {53},
  doi      = {10.1108/17410400410569125},
  url      = {https://doi.org/10.1108/17410400410569125},
  urldate  = {2025-05-05},
}

@Article{Nan2022,
  author    = {Nan, Ying-Gang and Yazd, Nazila Safari and Chapalo, Ivan and Chah, Karima and Hu, Xuehao and Mégret, Patrice},
  journal   = {Sensors},
  title     = {Properties of {Fiber} {Bragg} {Grating} in {CYTOP} {Fiber} {Response} to {Temperature}, {Humidity}, and {Strain} {Using} {Factorial} {Design}},
  year      = {2022},
  issn      = {1424-8220},
  month     = jan,
  note      = {2 citations (Crossref) [2025-05-05] Number: 5 Publisher: Multidisciplinary Digital Publishing Institute},
  number    = {5},
  pages     = {1934},
  volume    = {22},
  copyright = {http://creativecommons.org/licenses/by/3.0/},
  doi       = {10.3390/s22051934},
  language  = {en},
  url       = {https://www.mdpi.com/1424-8220/22/5/1934},
  urldate   = {2025-05-05},
}

@Article{Sheshadri2021,
  author     = {Sheshadri, Rohith and Nagaraj, Mohan and Lakshmikanthan, Avinash and Chandrashekarappa, Manjunath Patel Gowdru and Pimenov, Danil Yu and Giasin, Khaled and Prasad, Raghupatruni Venkata Satya and Wojciechowski, Szymon},
  journal    = {Journal of Materials Research and Technology},
  title      = {Experimental investigation of selective laser melting parameters for higher surface quality and microhardness properties: taguchi and super ranking concept approaches},
  year       = {2021},
  issn       = {2238-7854},
  month      = sep,
  note       = {40 citations (Crossref) [2025-05-05]},
  pages      = {2586--2600},
  volume     = {14},
  doi        = {10.1016/j.jmrt.2021.07.144},
  shorttitle = {Experimental investigation of selective laser melting parameters for higher surface quality and microhardness properties},
  url        = {https://www.sciencedirect.com/science/article/pii/S2238785421008073},
  urldate    = {2025-05-05},
}

@Article{Canel2012,
  author   = {Canel, Timur and Kaya, A. Uğur and Çelik, Bekir},
  journal  = {Optics \& Laser Technology},
  title    = {Parameter optimization of nanosecond laser for microdrilling on {PVC} by {Taguchi} method},
  year     = {2012},
  issn     = {0030-3992},
  month    = nov,
  note     = {60 citations (Crossref) [2025-05-05]},
  number   = {8},
  pages    = {2347--2353},
  volume   = {44},
  doi      = {10.1016/j.optlastec.2012.04.023},
  url      = {https://www.sciencedirect.com/science/article/pii/S0030399212001843},
  urldate  = {2025-05-05},
}

@Article{Weng2007,
  author     = {Weng, Wei-Chung and Yang, Fan and Elsherbeni, Atef Z.},
  journal    = {IEEE Transactions on Antennas and Propagation},
  title      = {Linear {Antenna} {Array} {Synthesis} {Using} {Taguchi}'s {Method}: {A} {Novel} {Optimization} {Technique} in {Electromagnetics}},
  year       = {2007},
  issn       = {1558-2221},
  month      = mar,
  note       = {154 citations (Crossref) [2025-05-05]},
  number     = {3},
  pages      = {723--730},
  volume     = {55},
  doi        = {10.1109/TAP.2007.891548},
  shorttitle = {Linear {Antenna} {Array} {Synthesis} {Using} {Taguchi}'s {Method}},
  url        = {https://ieeexplore.ieee.org/document/4120302},
  urldate    = {2025-05-05},
}

@Article{Mazoukh2024,
  author   = {Mazoukh, Celine and Di Lauro, Luigi and Alamgir, Imtiaz and Fischer, Bennet and Perron, Nicolas and Aadhi, A. and Eshaghi, Armaghan and Little, Brent E. and Chu, Sai T. and Moss, David J. and Morandotti, Roberto},
  journal  = {Communications Physics},
  title    = {Genetic algorithm-enhanced microcomb state generation},
  year     = {2024},
  issn     = {2399-3650},
  month    = mar,
  note     = {10 citations (Crossref) [2025-05-05]},
  number   = {1},
  pages    = {81},
  volume   = {7},
  doi      = {10.1038/s42005-024-01558-0},
  url      = {https://doi.org/10.1038/s42005-024-01558-0},
}

@Article{Lapre2023,
  author   = {Lapre, Coraline and Meng, Fanchao and Hary, Mathilde and Finot, Christophe and Genty, Goëry and Dudley, John M.},
  journal  = {Scientific Reports},
  title    = {Genetic algorithm optimization of broadband operation in a noise-like pulse fiber laser},
  year     = {2023},
  issn     = {2045-2322},
  month    = feb,
  note     = {12 citations (Crossref) [2025-05-05]},
  number   = {1},
  pages    = {1865},
  volume   = {13},
  doi      = {10.1038/s41598-023-28689-8},
  url      = {https://doi.org/10.1038/s41598-023-28689-8},
}

@InCollection{Taguchi1989,
  author    = {Taguchi, G. and Phadke, M. S.},
  booktitle = {Quality {Control}, {Robust} {Design}, and the {Taguchi} {Method}},
  publisher = {Springer US},
  title     = {Quality {Engineering} through {Design} {Optimization}},
  year      = {1989},
  address   = {Boston, MA},
  editor    = {Dehnad, Khosrow},
  isbn      = {978-1-4684-1472-1},
  pages     = {77--96},
  doi       = {10.1007/978-1-4684-1472-1_5},
  url       = {https://doi.org/10.1007/978-1-4684-1472-1_5},
}

@Article{MAZ2016,
  author  = {{Mohammad Asif Zaman}},
  journal = {Optical Engineering},
  title   = {Application of {Taguchi}’s method to optimize fiber {Raman} amplifier},
  year    = {2016},
  month   = apr,
  note    = {4 citations (Crossref) [2025-05-05]},
  number  = {4},
  pages   = {046103},
  volume  = {55},
  doi     = {10.1117/1.OE.55.4.046103},
  url     = {https://doi.org/10.1117/1.OE.55.4.046103},
}

@Article{Blow1991,
  author   = {Blow, K.J. and Doran, N.J.},
  journal  = {IEEE Photonics Technology Letters},
  title    = {Average soliton dynamics and the operation of soliton systems with lumped amplifiers},
  year     = {1991},
  issn     = {1941-0174},
  month    = apr,
  note     = {145 citations (Crossref) [2025-05-05]},
  number   = {4},
  pages    = {369--371},
  volume   = {3},
  doi      = {10.1109/68.82115},
  url      = {https://ieeexplore.ieee.org/document/82115},
  urldate  = {2025-05-05},
}

@Article{Hasegawa1990,
  author    = {Hasegawa, Akira and Kodama, Yuji},
  journal   = {Optics Letters},
  title     = {Guiding-center soliton in optical fibers},
  year      = {1990},
  issn      = {1539-4794},
  month     = dec,
  note      = {301 citations (Crossref) [2025-05-05] Publisher: Optica Publishing Group},
  number    = {24},
  pages     = {1443--1445},
  volume    = {15},
  copyright = {© 1990 Optical Society of America},
  doi       = {10.1364/OL.15.001443},
  language  = {EN},
  url       = {https://opg.optica.org/ol/abstract.cfm?uri=ol-15-24-1443},
  urldate   = {2025-05-05},
}

@Article{Tajima1987,
  author    = {Tajima, Kazuhito},
  journal   = {Optics Letters},
  title     = {Compensation of soliton broadening in nonlinear optical fibers with loss},
  year      = {1987},
  issn      = {1539-4794},
  month     = jan,
  note      = {183 citations (Crossref) [2025-05-05] Publisher: Optica Publishing Group},
  number    = {1},
  pages     = {54--56},
  volume    = {12},
  copyright = {© 1987 Optical Society of America},
  doi       = {10.1364/OL.12.000054},
  language  = {EN},
  url       = {https://opg.optica.org/ol/abstract.cfm?uri=ol-12-1-54},
  urldate   = {2025-05-05},
}

@Article{Richardson1995,
  author  = {{Richardson D.J.} and {Chamberlin R.P.} and {Dong L.} and {Payne D.N.}},
  journal = {Electronics Letters},
  title   = {High quality soliton loss-compensation in 38 km dispersion-decreasing fibre},
  year    = {1995},
  month   = sep,
  note    = {33 citations (Crossref) [2025-05-05] Publisher: The Institution of Engineering and Technology},
  number  = {19},
  pages   = {1681--1682},
  volume  = {31},
  doi     = {10.1049/el:19951144},
  url     = {https://doi.org/10.1049/el:19951144},
  urldate = {2025-05-05},
}

@InCollection{Hedayat1999,
  author       = {Hedayat, A. S. and Sloane, N. J. A. and Stufken, John},
  booktitle    = {Orthogonal {Arrays}},
  publisher    = {Springer New York},
  title        = {Orthogonal {Arrays} and {Hadamard} {Matrices}},
  year         = {1999},
  address      = {New York, NY},
  isbn         = {978-1-4612-7158-1 978-1-4612-1478-6},
  note         = {Series Title: Springer Series in Statistics},
  pages        = {145--166},
  collaborator = {Hedayat, A. S. and Sloane, N. J. A. and Stufken, John},
  doi          = {10.1007/978-1-4612-1478-6_7},
  language     = {en},
  url          = {http://link.springer.com/10.1007/978-1-4612-1478-6_7},
  urldate      = {2025-03-22},
}

@Misc{Sloane,
  author    = {Neil J. A. Sloane},
  title     = {A Library of Orthogonal Arrays},
  url       = {http://neilsloane.com/oadir/},
}

@Book{Ross1988,
  author    = {Phillip J Ross},
  publisher = {McGraw-Hill Book Company},
  title     = {Taguchi techniques for quality engineering : loss function, orthogonal experiments, parameter and tolerance design},
  year      = {1988},
}

\end{backmatter}
\newpage

\section*{Supplementary information}
Figures S.1 and S.2 shows a comparison of performances of the Taguchi, genetic algorithm (GA) and particle swarm optimization (PSO) routines for the guiding center soliton problem from Sec. 3.1 of the main paper. The results here are shown for $N_0  = 1.3285$, i.e. $N_Theory$. 
For the Taguchi method RR is chosen to be 0.8, which promotes exploration over exploitation. To ensure a fair comparison between the methods, the parameters for the GA and PSO routines are also chosen to promote exploration. The table below shows the list of parameters that were used for the two routines. The rest of the simulation parameters and the definition of the objective function have been kept identical. 

\begin{table}
\caption*{Table S.1 - Parameters chosen for the GA and PSO routines}
\begin{center}
\begin{tabular}{ll}
\toprule
\textbf{Genetic Algorithm (GA)} & \textbf{Particle Swarm Optimization (PSO)} \\
\midrule
Number of individuals – 9 & Swarm size – 9 \\
Maximum number of generations – 100 & Maximum iterations – 100 \\
Selection function – Tournament & Inertia range – [0.9 1.1] \\
Crossover function – 0.2 & Self-adjustment weight – 2.5 \\
Elite individuals – 1 & Social adjustment weight – 1.0\\
 & Minimum neighbours fraction – 0.25\\
\bottomrule
\end{tabular}
\end{center}
\end{table}

\begin{figure}[h]
  \centering
  \includegraphics[width=\linewidth]{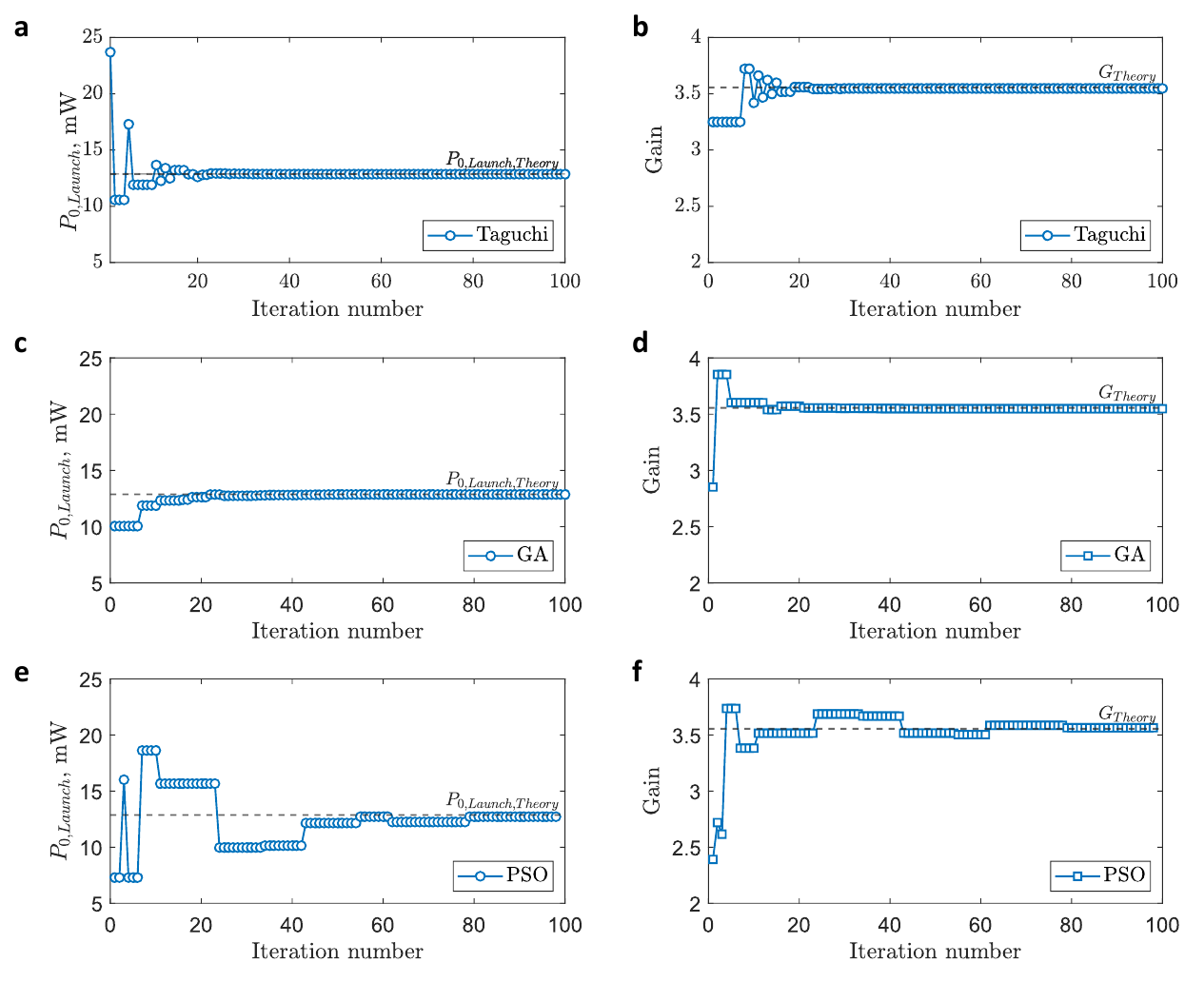}
\caption*{\textbf{Fig. S.1} Comparison between the Taguchi, GA and PSO routines.}
\end{figure}

\begin{figure}[h]
  \centering
  \includegraphics[width=\linewidth]{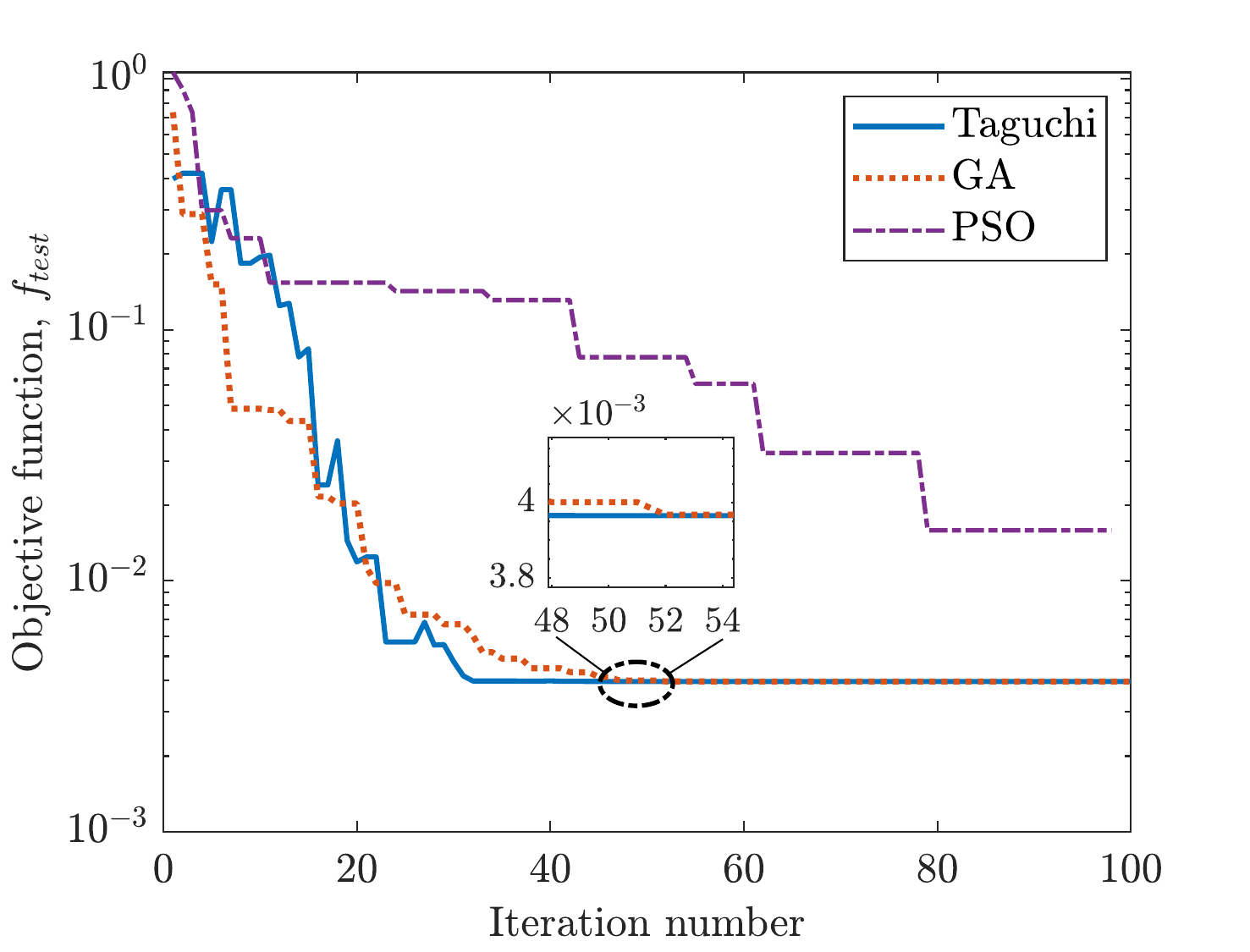}
\caption*{\textbf{Fig. S.2} Comparison of convergence of the objective function $f_test$ between the Taguchi, GA and PSO routines.}
\end{figure}

In both cases above, the number of individuals/swarm size is taken to be 9 to match with the number of experimental runs (or equivalently, combination of factors) of the Taguchi method. For the GA, the crossover function is set at 0.2 for promoting exploration. The number of elite individuals carried over to the next generation without any change in genes is 1, which again is set to match the update step of the Taguchi method. For the PSO, a higher value is assigned to the self-adjustment weight over the social adjustment weight, which promotes exploration. Similarly, the individual particles in the swarm interact only with a small fraction of the neighbourhood, which also promotes exploration. The same applies to the inertia range, which ensures that the particles do not slow down quickly, nor are they influenced by personal or social best outcomes. As the number of experimental runs/genes/swarm size are kept the same across the methods along with the adaptive split step propagation algorithm, the duration of execution of the respective routines are fairly similar and being primarily determined by the time of execution of the adaptive split-step algorithm. This is not necessary in the case of the Taguchi method, as it is of a deterministic nature owing to its use of the orthogonal arrays. 

Figure S.1 shows the convergence characteristics of the GA and PSO routines in comparison to the Taguchi method. The figures qualitatively indicate that the Taguchi method and the GA converge appreciably faster than the PSO. Fig S.2 presents a more quantitative perspective on the nature of convergence by plotting the value of the objective function $f_test$ over iterations. This shows that the slope of convergence of the Taguchi method is faster than the GA. Furthermore, while the GA takes about 50 iterations to reach a minimum value of $f_test$, the Taguchi method is able to do the same in about 30 iterations. Note that the $\sim$20 extra iterations correspond to 180 split-step propagations for the guiding center soliton problem resulting in a computational overhead. Thus in this context too the Taguchi method potentially offers a significant speed up.

While further systematic analysis is required to explore and compare the convergence characteristics across optimization techniques, the results presented here show that in light of its simplicity of implementation and interpretation, the Taguchi method can be used very effectively in nonlinear pulse propagation problems.

\end{document}